\documentclass[superscriptaddress,twocolumn,prb]{revtex4}
\usepackage{amssymb}
\usepackage{color}

\newcommand{\cV}{\mathcal{V}}
\newcommand{\cP}{\mathcal{P}}
\newcommand{\cL}{\mathcal{L}}
\newcommand{\cM}{\mathcal{M}}
\newcommand{\cN}{\mathcal{N}}

\begin{document}

\title{$1/f$ noise in hopping conduction: Role of multi-site aggregates}
\author{V. I. Kozub}
\affiliation{A. F. Ioffe  Physico-Technical Institute of Russian
Academy of Sciences, 194021 St. Petersburg, Russia}
\affiliation{Argonne National Laboratory, 9700 S. Cass Av.,
Argonne, IL 60439, USA}
  \author{Y. M. Galperin}
\affiliation{Department of Physics, University of Oslo, PO Box
1048 Blindern, 0316 Oslo, Norway} \affiliation{A. F. Ioffe
Physico-Technical Institute of Russian Academy of Sciences, 194021
St. Petersburg, Russia} \affiliation{Argonne National Laboratory,
9700 S. Cass Av., Argonne, IL 60439, USA}
\author{V. Vinokur}
\affiliation{Argonne National Laboratory, 9700 S. Cass Av.,
Argonne, IL 60439, USA} \pacs{71.23.Cq, 72.70.+m, 72.20.Ee,
72.80.Sk}

\begin{abstract}
We propose a mechanism for $1/f$-type noise in hopping insulators
based on the multi-electron charge redistribution within the
specific aggregates of the localized states located in the
vicinity of the critical resistors. We predict that the noise with
$1/f$-type spectrum extends down to practically arbitrarily low
frequencies.
\end{abstract}

\maketitle


A nature of the underlying mechanism of $1/f$ low-frequency noise
spectrum in hopping conductors is a part of the more general
problem of the universal origin of $1/f$, and had been an
unresolved issue for decades.~\cite{1,2,3,4,6,7,8,9,11,12} A
general intrinsic reason for the $1/f$ spectrum is the
exponentially wide distribution of the relevant relaxation times
in the system.  Thus the issue of the mechanism of $1/f$ spectrum
reduces to the question of the microscopic origin of such a
distribution. The low temperature transport in doped
semiconductors occurs via electron hopping between the donors (or
hole hopping between acceptors when considering $p$-type
semiconductors).  The hopping distance is determined by the
balance between the tunneling probability over this distance and
the probability of the thermal activation necessary to accomodate
the difference between the initial and final levels.  Optimization
of these probabilities cuts the effectively conducting band around
the Fermi level. This hopping mechanism is called variable range
hopping (VRH) transport. With the decreasing temperature this
effective impurity band gets narrower and the electron has to
tunnel over a larger distance to find the appropriate destination
state, since the levels at the close by donors are separated by a
large gap, see Ref.~\onlinecite{Shklovskii-Efros} and references
therein. In this regime the conducting paths form a dilute
percolation cluster, and the system resistance is controlled by
the very rare {\textit critical hopping resistors $R_c$}.  The
conductivity of the critical resistors are most susceptible to the
dynamic fluctuations of charge distribution in a body of the
conductor and they become the source of the current noise.

The crucial question associated with the origin of the noise is
whether the $1/f$ law goes indefinitely to low frequencies and if
not, what is the nature of its lower bound. This problem was first
addressed by Shklovskii~\cite{7} and Kogan\&Shklovskii~\cite{8}
who attributed the origin of the noise to the charge traps
affecting the electron concentration on the percolation cluster
that supports the current in the conductor. This model predicts
the saturation of the noise spectrum at low frequencies.  The
reason for the saturation is that no trap can be perfectly
isolated from the rest of the hopping sites since their spatial
distribution is Poissonian. Consequently, to secure a very large
relaxation time one has to find a very rare trap specially
isolated from the other hopping sites. These ideas of
Refs.~\onlinecite{7,8} were elaborated in the recent paper by
Shklovskii~\cite{12} devoted to the temperature dependence of the
noise spectrum.  The alternative idea was put forward by
Kogan~\cite{11} who assumed that a hopping system is in fact the
Coulomb glass and can therefore  have arbitrarily low intrinsic
frequencies due to transitions between the different metastable
``pseudo-ground" states separated by unlimitedly large barriers.
Accordingly, there is no saturation in the noise spectrum down to
extremely low frequencies. Unfortunately, no analytical
description for this mechanism was proposed.

Here we revisit the problem of the low-frequency noise in hopping
conductors and develop the approach for calculating the noise
spectrum based on the idea of multi-site bistable
\textit{aggregates}. While switching between their pseudo-ground
states, the aggregates produce low-frequency electrical noise,
which is ``read-out" by the sites belonging to the backbone
percolation cluster. The read-out mechanism suggested
earlier~\cite{9} attributes the noise to fluctuations of the
effective resistors forming bonds of the percolation cluster with
correlation length $\cL$. The fluctuations of each resistor are
due to the noise produced by the nearest bistable aggregate. Since
the switching times of these aggregates are spread in an
exponentially broad interval the noise spectrum turns out to be of
$1/f$ type,

In the original approach~\cite{9} the aggregates consisted of
pairs of sites. However, the probability of finding such pairs -
\textit{fluctuators} - is limited by the fact that such fluctuator
be located in a ``pore" free of other sites able to facilitate
electron hopping.~\cite{12} Consequently, the maximal switching
time are limited that leads to a cut-off in the noise spectrum
Here we generalize the approach~\cite{9} including into the
consideration \textit{multi-site aggregates} that can assume two
distinct charge configurations with the nearly same energy. In
this respect, the aggregates that can be viewed as two-level
systems.  It is essential that transitions between these states
occur via the multi-electron hops, either coherent or incoherent.
The probabilities of such transitions are strongly suppressed
comparing to the case of isolated pairs, and this is the source of
very large intrinsic switching times.
Consequently, the cut-off frequency can be extremely low. The
suggested model of multi-site bistable aggregates is a ``bridge"
between the ideas of small fluctuators~\cite{9,12} and of
pseudo-ground state of the whole system.~\cite{11}

\section{Model}

To quantify the concept of an aggregate let us start from the
simplest one -- a fluctuator consisting of the two trapping sites
occupied by a single electron. Such pairs have two levels with
small energy separation $E \lesssim T$ and their relaxation time,
$\tau_p$, is related to the tunneling between the states;
%
$\tau_p^{-1}=\nu_0\, e^{-2 r_p/a}$ where $\nu$ is some (generally
temperature-dependent) pre-exponential factor, $r_p$ is the
spatial separation between the sites while $a$ is the localization
length. The values of $\tau$ should be much larger than the
typical hopping times, $\tau_h=\nu_0^{-1}e^{2r_h/a}$, along the
current-carrying percolation cluster, therefore $r_p$ should be
larger than the typical hopping distance in the cluster,
$r_h=a\xi$. Here $\xi$ is the connectivity parameter of the
percolation theory entering the exponent of critical hopping
resistor, $R_c=R_0\, e^\xi$, see
Ref.~\onlinecite{Shklovskii-Efros} and references therein. We
consider an isolated pair, therefore there is no any other site in
a close neighborhood (i.~e. within the distance $\tilde{r} \ll
r_p$) from either of the centers, that could facilitate the
processes with the rates $\gg \tau_p^{-1}$.
One can express this requirement as the inequality
\begin{equation}\label{pair}
\frac{2r_p}{a} \leq \frac{\Delta}{T} + \frac{2 {\tilde r}}{a}=\xi
\end{equation}
where $\Delta$ is the relevant energy band width. To relate this
bandwidth to the distance $\tilde{r}$ we estimate $\tilde{r}$ as
$\tilde{r}=(3/4\pi n)^{1/3}$ where $n$ is the concentration of the
relevant centers. Since for a Coulomb-gap controlled system in the
VRH regime
$n=(2/3)(\kappa \Delta/e^2)^3$ we obtain $\Delta \simeq
(9/8\pi)^{1/3}(e^2/\kappa \tilde {r})$. Optimizing the r.h.s. of
Eq.~(\ref{pair}) with respect to  $\tilde r$ we find
$\tau_p \lesssim \tau_h e^{ \xi ( \sqrt{2} - 1)}$.
Here $\xi=(T_0/T)^{1/2}$,
$T_0=e^2/{\kappa}a$.~\cite{Shklovskii-Efros}

Now we consider the aggregates consisting of 4 hopping sites
occupied by 2 electrons. These close to square arrays are the
simplest \textit{next hierarchy level} aggregates realizing the
two level system (the three-site aggregates are reduced to pairs,
since even a slight deviation from the perfect arrangement lifts
the triple degeneracy and gives rise to an exponentially wide gap
separating the split levels of the closest pair from the third
level).
A degenerate configurations requiring the least energy arise when
two electrons are trapped at the diagonally opposite sites.
Labelling the sites as $1$, $2$, $3$ and $4$, we say that the
pairs are formed by the occupied sites $1 \leftrightarrow 2$, and
$3 \leftrightarrow 4$, respectively. For the configuration where
sites $2$ and $4$ are occupied one has
\begin{eqnarray}\label{1}
\varepsilon_2 + \frac{e^2}{\kappa r_{42}} < \mu \, , \hskip 0.5cm
\varepsilon_1 + \frac{e^2}{\kappa r_{12}} + \frac{e^2}{\kappa
r_{14}} > \mu \, ,
 \nonumber\\
\varepsilon_4 + \frac{e^2}{\kappa r_{42}} < \mu \, , \hskip 0.5cm
\varepsilon_3 + \frac{e^2}{\kappa r_{43}} + \frac{e^2}{\kappa
r_{23}} > \mu \, ;
\end{eqnarray}
the configuration where sites $1$ and $3$ are occupied obeys:
\begin{eqnarray}\label{2}
\varepsilon_1 + \frac{e^2}{\kappa r_{13}} < \mu\, , \hskip 0.5cm
\varepsilon_2 + \frac{e^2}{\kappa r_{12}} + \frac{e^2}{\kappa
r_{23}} > \mu \, ,
 \nonumber\\
\varepsilon_3 + \frac{e^2}{\kappa r_{13}} < \mu \, , \hskip 0.5cm
\varepsilon_4 + \frac{e^2}{\kappa r_{4,1}} + \frac{e^2}{\kappa
r_{4,3}} > \mu \, .
\end{eqnarray}
Here $\varepsilon_i$ are the single-particle energies of the
sites, $r_{ij}$ are the sites spacings, and $\mu$ is the chemical
potential. The energy splitting between the two configurations is
\begin{equation}
E \equiv E_1 - E_2 = 2\frac{e^2}{\kappa}\left(\frac{1}{r_{24}} -
  \frac{1}{r_{13}}\right) +
\varepsilon_2 + \varepsilon_4 - \varepsilon_1 - \varepsilon_3 \, .
\end{equation}
The relaxation processes associated with the transitions between
the two configurations with the almost equal energies are the
process that involve the simultaneous exchange of electrons
between the sites $1 \to 2$ and $3 \to 4$.  All other processes
increase the Coulomb energy and therefore are unlikely to occur.
 The switch can be either due to a
multi-electron (a two-electron in our case) tunneling first
suggested by Pollak~\cite{Pollak} or due to incoherent processes.

 As follows from the above mentioned correlation, the aggregate
formed by the sites with large single-particle energies would have
short intersite distances and, accordingly, high rates of the
multi-electron hops. Since the slow relaxing aggregates should
have relatively large larger intersite separations, their
single-particle energies should be relatively small.

To  proceed further it is convenient to map the aggregates onto a
spin system.  A single pair is mapped on the $1/2$-spin (the spin
direction is taken from the occupied to the unoccupied site), the
four-site aggregates are represented as a pair of the interacting
antiparallel spins. The interaction between the spins forming an
aggregate is antiferromagnetic since the Coulomb repulsion pushes
electrons belonging to the adjacent pairs apart. Adding more pairs
we map larger aggregates to clusters of $3, 4, ... N$ spins. This
model is a \textit{minimal} model for the general Coulomb glass
since we allow only for the transitions that are represented by
flips of spins. In the absence of randomness the system is
antiferromagnet, disorder may, in principal, drive it to the spin
glass.

The $2N$-site aggregates (or antiferromagnet $N$-spin clusters)
must obey restrictions similar to those given by Eqs.~(\ref{1})
and (\ref{2}). To estimate the density of states for the
aggregates we shall take into account that in course of the
reversal of the ``spins" the energy of each site crosses the Fermi
level. Thus we should compare the change of the site energy
resulting from the aggregate rearrangement and the ``static"
scatter of the site energies.

To do so, let us specify some energy $\varepsilon_f$ within the
Coulomb gap and consider the sites with energy band $\varepsilon_i
< \varepsilon_f$. Let us consider the aggregates formed from these
sites in the same way as it was considered above. For the site
energy we have, cf. with Ref.~\onlinecite{Shklovskii-Efros},
\begin{equation}\label{ag}
\varepsilon_i = \varphi_i + V_i, \quad V_i \equiv
\frac{e^2}{\kappa}\sum_{j \neq i} \frac{(1-n_j)}{r_{ij}}\, .
\end{equation}
Here $\varphi_i$ is the potential induced on the site $i$ by the
background charges not included in the aggregate;  the sum in
$V_i$ is calculated over the sites forming the aggregate. Thus
$\varphi_i$ is due to the disorder while $V_i$ depends on the
state of the aggregate.

 To estimate $V_i$ let us split it into the parts having different
 symmetry with respect to rearrangement of the aggregate between
 its two metastable states: $V_i=(1/2)(V_i^+ + V_i^-  s_i)$. Here
 $s_i =\pm 1$. The symmetric part can be treated as a contribution to the
 potential $\varphi_i$.
Assuming the potentials $\varphi_i$ and $V_i^{\pm}$ to be random
and uncorrelated, we write
\begin{equation}\label{random}
\overline{\varepsilon_i^2} = \overline{\varphi_i^2} + \frac{1}{4}
\overline{ (V_i^+)^2} + \frac{1}{4}\overline{ (V_i^-)^2}\, .
\end{equation}
Since we are interested in the states within the Coulomb gap, the
density of states for the single-particle energies $\varepsilon_i$
is given by the Efros-Shklovskii law, $g \propto \varepsilon_i^2$.
Thus the typical distance between the sites as
\begin{equation}
\bar{r} = \left[2 \int_0^{\varepsilon_f}g(\varepsilon){\rm
d}\varepsilon\right]^{-1/3}=(3/2)^{1/3} e^2/\kappa \varepsilon_f\,
.
\end{equation}
One expects that the potential formed on the site $i$ by the other
sites forming the aggregate can estimated as $\bar{V}_\pm\equiv
\left( \overline{(V_i^\pm)^2}\right)^{1/2} = \zeta_\pm
\varepsilon_f $ where $\zeta_\pm \sim 1$ are numerical parameters
depending on the aggregate geometry. Making use of
Eq.~(\ref{random}) one has
\begin{equation}
\overline{\varphi^2} = \varepsilon_f^2(1 - \zeta^2_+/4
-\zeta_-^2/4)\, .
\end{equation}
Since the typical variation of the site energy, $\delta V \equiv
\bar{V}_-=\zeta_-\varepsilon_f$ we obtain the following estimates
\begin{equation}\label{bardeltav}
\bar{V}_-=\zeta_-\varepsilon_f, \ \bar{V}_+=\zeta_+ \varepsilon_f,
\ \bar{\varphi}=\varepsilon_f \sqrt{1-\zeta_-^2/4 -\zeta_+^2/4}\,
.
\end{equation}
Note that the relations between $\bar{V}_-$, $\bar{V}_+$ and
$\bar{\phi}$ do not depend on $\varepsilon_f$ provided we deal
with the states within the Coulomb gap. One can expect that these
ratios are of the order of unity.

Basing on the discussion given above let us estimate
the aggregates density of states. First, let us note that for all
the sites belonging to the aggregate the inequality $V_i^-
> V_i^+ + 2\varphi_i$ must be met. Indeed, only such states
cross the Fermi level
The probability to form an aggregate depends on two parameters -
the ratios $\zeta_-/\zeta_+$ and $ \zeta_-/2\varphi$. Assuming
that $\zeta_+/\zeta_- \lesssim 1$ and $2\bar{\varphi}\gtrsim
\zeta_-$  one concludes that to form an aggregate
one should deal with the sites with the potentials $\varphi_i$
less than their average scatter. Since the Coulomb gap exists only
for the single-electron energies, $\varepsilon_i$ rather than for
disorder-induced potential $\varphi$, the distribution of the
random potential $\varphi$ , $\cP(\varphi)$, is smooth for small
$\varphi$ and one can put $\cP(\varphi)\approx 1/2\bar{\varphi}$.
Consequently, he relative weight of the aggregate sites is in this
case given by
$$\frac{1}{2\bar{\varphi}} \int^{(\bar{V}_- - \bar{V}_+)/2}_0
\cP(\varphi) \, {\rm d} \varphi =
\frac{\zeta_--\zeta_+}{\sqrt{4-\zeta_-^2 -\zeta_+^2}} \equiv
\lambda \, . $$ These considerations are valid until $\lambda <1$.
 In this case relative weight of the aggregate formed by $2N$ sites
 is $\lambda^{2N}=e^{-2N\ln1/\lambda}$.
Thus at $\lambda <1$ is the probability to form a metastable
aggregate is exponentially small in terms of the number of the
sites involved in the aggregate.  This will be mainly discussed in
what follows.

However, there is no fundamental principle requiring $\lambda \le
1$.~\cite{endnote} This is why later we will also discuss the
scenarios where the multistable aggregates can be realized with a
probability close to unity.

\section{The distribution of the relaxation rates.}\label{s1}

To calculate noise in the system one needs  density of states, $P
(N,r,E)$, of finding the aggregate with $N$ sites and distance $r$
between the sites per unit energy interval, $E$. This density can
be expressed in  the form
\begin{eqnarray}
  \label{eq:006}
  P(N,r,E)&=&\frac{\lambda^N}{\sqrt{\pi N} N r^3\cdot e^2/\kappa r}
\nonumber \\
  &=& \frac{2.7 \lambda^N}{T_{ES}\sqrt{\pi} N^{3/2}ar^2} \, .
\end{eqnarray}
Here $Nr^3$ is the volume of the aggregate while
$\sqrt{N}e^2/\kappa r$ is the energy bandwidth for small energies.

Let us estimate the distribution function, $\cP(\nu)$, of the
relaxation rates, $\nu$, for the aggregate rearrangement. We
define the rate as
\begin{equation}
  \label{eq:001}
  \nu=\nu_0\max\left\{e^{-N^{2/3}\xi^2/2.7\rho},e^{-2N\rho}\right\}\, .
\end{equation}
Here $\rho \equiv r/a$, $\xi \equiv (T_{ES}/T)^{1/2}$. The first
term in parentheses describes formation of a ``domain wall" in the
aggregate and the second term corresponds to coherent tunneling
transitions leading to re-charging of all aggregate sites.

To calculate the distribution of switching rates let us introduce
the function $L(N,\rho)$ as
\begin{equation}
  \label{eq:002}
  L(N,\rho)=-\ln \max\left\{e^{-N^{2/3}\xi^2/2.7\rho},e^{-2N\rho}\right\}\, .
\end{equation}
Thus $L(N,\rho)=\ln(\nu_0/\nu) \equiv \cM$. Denoting $\rho_N$ by
the equality $N^{2/3}\xi^2/2.7\rho_N=2N\rho_N$,
\begin{equation}
  \label{eq:003}
\rho_N=0.43\xi/N^{1/6}\, ,
\end{equation}
one can express $L(N,\rho)$ as
$$L(N,\rho)=-\ln \max\left\{e^{-0.86N^{5/6}\xi
    (\rho_N/\rho)},e^{-0.86N^{5/6}\xi (\rho/\rho_N)} \right\}\, .$$
Let us now invert the above dependence. Since
\begin{equation}\label{ldep}
L(N,\rho) \approx \left\{\begin{array} {ll} 0.86N^{5/6}\xi
(\rho/\rho_N)\, , & \rho \gg \rho_N \\
0.86N^{5/6}\xi
    (\rho_N/\rho)\, , & \rho \gg \rho_N\, . \end{array} \right.
    \end{equation}
the inverted function, $\rho(L)$, has two branches:
\begin{equation}
  \label{eq:004}
  \frac{\rho(L)}{\rho_N}=\left\{ \begin{array}{l}
 (L/L_N)\\
(L_N/L)\, ,
\end{array}
\right. \quad \left|\frac{\partial \rho}{\partial
L}\right|=\frac{\rho_N}{L_N}\left\{
  \begin{array}{l} 1,\\(L_N/L)^2
  \end{array} \right. \, .
\end{equation}
Here  $L_N\equiv 0.86N^{5/6}\xi$, thus $\rho_N/L_N=1/2N$. Note
that $L(N,\rho) \le L_N$. Now we are ready to calculate the
distribution function
\begin{eqnarray}
  \label{eq:005}
  \cP(\nu,E)&=&4\pi \int dN\, r^2 dr\, P(N,r,E)\, \delta \left(\nu -\nu_0\,
  e^{-L(r,N)} \right) \nonumber  \\
&=& \frac{4\pi a^2}{\nu}\int dN\,\rho^2 d\rho \, P(N,\rho,E)\,
\delta \left[\rho-\rho(L)
  \right]\left|\frac{\partial \rho}{\partial L}\right|
\, . \nonumber
\end{eqnarray}
Substituting $P(N,r,E)$ from Eq.~(\ref{eq:006}) we obtain
\begin{equation}
 \label{eq:009}
 \cP(\nu)=\frac{1}{\nu}\,
  \frac{2\pi\cdot2.7}{T_{ES}\sqrt{\pi}}\int_{N_c}^\infty
  \frac{ dN\,  e^{-\gamma N}}{N^{5/2}}
  \,\left[1+\left(\frac{L_N}{\cM}\right)^2 \right]
 \end{equation}
  where $\gamma \equiv \ln 1/\lambda$, while
  $N_c(\cM)=(\cM/0.86\xi)^{6/5}$.
Substituting these values to Eq.~(\ref{eq:009}) we get,
\begin{equation}
\label{eq:012} \cP(\nu) \approx\frac{1}{\nu}\,
  \frac{1}{T_{ES}}\frac{4\sqrt{\pi}\cdot 2.7}{\gamma
  N_c^{5/2}}e^{-\gamma N_c}\, .
\end{equation}
In this approximation the density of states $\cP$ does not depend
on the energy splitting $E$. The expression (\ref{eq:012}) is
valid at $\gamma N_c =\left[(\gamma^{5/6}/0.86 \xi)
\ln(\nu_0/\nu)\right]^{6/5} \gtrsim 1$, or
\begin{equation} \label{ineq02}
\ln (\nu_0/ \nu) \gtrsim \xi/\gamma^{5/6}\, .
\end{equation}
The product $\gamma N_c$ can be expressed as
$\ln(\nu_0/\nu)^\alpha$ where
$$ \alpha (\nu)=\frac{\gamma^{5/6}}{0.86 \xi}
\left[\frac {\gamma^{5/6}}{0.86 \xi}
\ln\left(\frac{\nu_0}{\nu}\right)\right]^{1/5}\, .$$ {}From
realistic estimates one can expect that the second factor does not
significantly change within the experimentally accessible
frequency interval. For example, assuming $\nu_0 = 10^{10}$ Hz we
get $\cM^{1/5}=1.87$ at $\nu =1$ Hz and $\cM^{1/5}=2.06$ at $\nu
=10^{-6}$ Hz. Consequently, for reasonable experimental conditions
the quantity $\alpha (\nu)$ can be replaced by
$\bar{\alpha}=\alpha(\bar{\nu})$ where $\bar{\nu}$ is some
characteristic frequency within the measurement interval. As
follows from the above estimate, for any feasible frequency
$\bar{\alpha} \ll 1$ and the noise spectrum is of the $1/f$ type.

As a result, the distribution of the relaxation rates can be
expressed as
\begin{equation}
  \label{eq:009a}
  \cP(\nu)\approx \frac{C}
  {\nu^{1-\bar{\alpha}}\ln^3(\nu_0/\nu)}\,
\frac{\nu_0^{-\bar{\alpha}}\xi^3}{T_{ES}} \, ,
  \end{equation}
  where $C \approx 12.2/\gamma$ is a numerical factor.
  The temperature dependence $\bar{\alpha} \propto \xi^{-6/5} \propto T^{3/5}$ can
  be used for verification of the proposed mechanism.

\section{Estimate of the noise intensity}

Switching between the aggregate configurations leads to the change
in the energies, $\delta \varepsilon_i^{(c)}$, of the sites
belonging to the percolation cluster and induces the current
noise. If $|\delta \varepsilon_i^{(c)}| \gtrsim T$ then the
fluctuations of the hopping resistor, $\rho_i$, are large,
$|\delta \rho_i| \sim \rho_i$ and can change the percolation
cluster structure  (see Ref.~\onlinecite{9}).

To be specific we consider an $N$-spin aggregate coupled to the
hopping resistor $\rho_i$ formed by the two hopping sites which we
label by indices 1 and 2. The fluctuations of the resistance are
estimated as
\begin{equation}
|\delta R_i/R_i| \sim \min \left(1,|\delta \varepsilon
_i|/T\right)\, , \quad \delta \varepsilon_i \equiv \delta
\varepsilon^{(c)}_{i,1} - \delta \varepsilon^{(c)}_{i,2}\, .
\end{equation}
If the distance between the resistor $i$ and the nearest
aggregate, $r_{i}$, is much larger than the typical distance
between the sites belonging to the hopping cluster, $r_{i} \gg
r_h$, the fluctuation in energy $\delta \varepsilon_i \approx
(\partial \delta \varepsilon^{(c)} /\partial r_{i}) r_h$ where
$\delta \varepsilon$ is the potential induced by the aggregate.
The latter can be estimated assuming that the total dipole moment
of the relevant aggregate is $er_c\sqrt{N_c}$ where $r_c\equiv a
\rho_{N_c}=0.43a\xi N_c^{-1/6}$. Consequently, $r_c=0.43 r_h
N_c^{-1/6}$ and
\begin{equation} \label{int1} \delta
\varepsilon^{(c)} \sim \frac{e^2 r_c \sqrt{N_c}}{\kappa r_{i}^2},
\quad \delta \varepsilon \sim \frac{e^2 r_h^2 N_c^{1/3}}{\kappa
r_{i}^3} \sim \frac{T_{ES}N_c^{1/3}\xi^2 a^3}{ r_{i}^3}\, .
\end{equation}

As we have already mentioned the noise is formed at the
exponentially rare critical resistors. Let us assume for a while
that the fluctuations of their resistances are small, $|\delta
R_i|/R_i \ll 1$.
In a linear approximation the total resistance fluctuation can be
written as
\begin{equation}
\frac{\delta R}{R} \sim \frac{1}{\cN} \sum_i \frac{\delta
R_i}{R_i}\, .
\end{equation}
Here $\cN$ is the number if critical resistors in the sample which
can be expressed through its volume, $\cV$, and the correlation
length of the percolation cluster, $\cL = r_h \xi = a\xi^2$, as
$\cN=\cV/\cL^3$. Assuming the aggregates acting upon different
critical resistor to be statistically independent we get.
\begin{equation}
\frac{\langle \delta R,\delta R\rangle_{\omega}}{R^2} \sim
\frac{1}{\cN^2}   \sum_i\frac{\langle \delta R_i, \delta
R_i\rangle_{\omega}}{R_i^2} \,.
\end{equation}

Now let us estimate of $\delta R_i$.  One has in mind that the
aggregates are also rarely distributed. This means that all the
critical resistors have at most one neighboring aggregate and the
fluctuation spectrum is:
$$
S(\omega) \equiv \frac{(\delta R_i)^2_{\omega}}{R_i^2}=v^2(r_i)\,
\frac{\nu_i}{\nu_i^2 + \omega ^2}\, \frac{1}{4\cosh^2 E_i /2T}\, .
$$ 
Here $\nu_i$ is the switching rate of the nearest aggregate with
energy splitting $E_i$, $v^2 (r_i) =\min \left[\left (\delta
\varepsilon(r_i)/T\right )^2,1\right]$ is the squared
dimensionless coupling. Since $\delta \varepsilon (r_i) \propto
r_i^{-3}$ there exists the specific distance, $r_T$, at which the
energy variation given by Eq.~(\ref{int1}) is equal to $T$:
\begin{equation} \label{rt}
r_T \approx a \xi^{4/3} N_c^{1/9}=r_h \xi^{1/3} N_c^{1/9} \ll
\cL\, .
\end{equation}
At $r_i \gg r_T$ the interaction strength $v^2(r_i)$ decays at
least as $\propto r_i^{-6}$. Consequently, only the nearest
aggregate is important.

Replacing summation over the critical resistors (and their nearest
aggregates) by averaging and integration over $E$ we obtain
\begin{equation}
S(\omega)  \sim \frac{\overline{v^2}T}{\cal N} \int_0^{\nu_0} {\rm
d} \nu\, \cP(\nu)\frac{\nu }{\nu^2 + \omega ^2}=
\frac{\pi\overline{ v^2}T}{2\cN}\cP(\omega)\, .
\end{equation}
Here $\overline {v^2} \equiv r_T^{-3}\int_{|\mathbf{r}|<r_T}
v^2(r)\, d^3r$, while the distribution function $\cP(\omega)$ is
given by Eq.~(\ref{eq:012}) or (\ref{eq:009a}). Using
Eq.~(\ref{int1}) we conclude that $\overline{v^2}$ is of the order
of unity, and
\begin{equation} \label{res1}
S(\omega) \sim \frac{\xi}{\cN}
\frac{\nu_0^{-\bar{\alpha}}}{\omega^{1-\bar{\alpha}}\ln^3(\nu_0/\omega)}
\, .
\end{equation}
The factor $\xi/\cN = \xi \cL^3/\cV = (a^3/\cV)\, \xi^7 \propto
T^{-7/2}$. Thus the the considered mechanism leads to a strongly
decreasing temperature dependence of the noise intensity. The
suggested procedure to estimate the noise intensity is valid if
each cell of the backbone cluster contains only one aggregate.
This implies the limitation to the cluster size
\begin{equation} \label{ineq01}
N_c r_c^2 \le \cL \quad \to \quad \ln(\nu_0/\omega) \le \xi^6\,.
\end{equation}
This requirement hold for any realistic frequency since $\xi \gg
1$. One can imagine another restriction relevant to the
percolation mechanism behind the resistivity
According to the $1/\omega$ spectrum, one expects that the
magnitude of the fluctuations increases with an increase of the
observation time. If the relative resistance fluctuation for any
relevant resistor,
$$\overline{\delta R_i^2}/R_i^2 = \cN
\overline{\delta R_i^2}/R^2\equiv (\cN/R^2) \int d\omega (\delta
R^2)_\omega $$ becomes comparable with unity, then one expects
that the percolation cluster is completely reconstructed by the
fluctuations. As a result, the fluctuations with lower frequencies
will not be observed in the sample resistance. As follows from
Eq.~(\ref{res1}), this requirement provides the frequency limit
for the validity of our calculation, $ \ln (\nu_0/\omega) \gg
\sqrt{\xi}$, which is automatically fulfilled for all the
frequencies less than the typical hopping frequency, $\nu_h=\nu_0
e^{-\xi}$.

\section{Coulomb glass scenarios}

The above considerations demonstrate many-site multistable
configurations in the hopping insulator. The probability of
finding such configurations depends on the interplay between the
Coulomb interaction and the random potential produced by stray
disorder, e.~g. by charged acceptors. The question which arises
here is whether there is a critical value of disorder
discriminating between the cases of exponentially low, $\propto
e^{-\gamma N}$, probability of finding such configurations and the
probability close to one. Earlier we relied upon rather strong
disorder assuming exponential decay of the probability with the
size of the metastable aggregate.

Now we will discuss the consequences of the intrinsically
correlated distribution of charges where there is no exponential
decay of the probability with the exponent proportional to the
number of sites in the aggregate. Namely, let us assume that
 there is a critical value $\gamma_c$ of the parameter $\gamma =\ln1 /\lambda$
such as that the systems with $\gamma < \gamma_c$ allow to form
the arbitrarily large metastable configurations. The independent
support of this idea (although again for model systems) is given
by theoretical considerations given in Ref.~\onlinecite{Ioffe}.
Namely, it is demonstrated that the sites within a Coulomb gap
demonstrate replica-symmetry breaking which can be related  to a
presence of extensive number of metastable states within a
thermodynamic limit $N \rightarrow \infty$. We also note that
multi-stable character of the ground states was also demonstrated
by several numerical simulations.~\cite{11,Laikhtman}

In principle, one can consider the following 3 scenarios:
\begin{itemize}
\item[(i)] \textit{Strong disorder}, $\gamma > \gamma_c$. Only
rare multistable aggregates are available due to the effect of
disorder, the probability exponentially decreasing with an
increase of the aggregate size.

\item[(ii)] \textit{Weak disorder}, $\gamma < \gamma_c$. The
multistable configurations are inherent for the system, which in
this case can be called the Coulomb glass. The results depends on
the typical distance of charge transfer:
\begin{itemize}
\item[a.] the charge transfer within the metastable aggregates is
restricted by neighboring sites. Correspondingly, the interactions
are dominated by the short range dipole-dipole forces (``dipolar
glass"),  and the short range ordering in the sites occupation
numbers still exists. In this case one expects that the local
glassy dynamics is not significantly affected by coupling with the
remote regions. Namely, the dynamics of a critical resistor is
dominated by the nearest aggregate with a specific relaxation
rate. One can expect that in this regime the system would have the
$1/\omega$ spectrum down to any practically achievable frequency,
and the exponent $\bar{\alpha}=0$.

\item[b.] The charge transfer at large distances is important. In
this case the local short-order configuration of the sites
occupation numbers can depend on the state of the remote regions
(``large-scale Coulomb glass"), and as a result a hierarchical
dynamics becomes possible. In this case the low-frequency noise
cannot be regarded as a superposition of statistically-independent
Poissonian telegraph-like fluctuations occurring in different
parts of the sample and acting on different backbone resistors.

Because of hierarchical dynamics of the aggregates one can expect
that fluctuation of \textit{each} site energy possesses $1/f$-type
spectrum. To the best of our knowledge, this situation has not
been analyzed yet in a convincing way.
\end{itemize}
The difference between the ``dipolar" and ``large-scale" Coulomb
glasses can be in principle revealed experimentally by analyzing
local occupation fluctuations by means of scanning tunnelling
spectroscopy. For the scenario (iia) the telegraph noise with a
given relaxation time is expected, while for (iib) a noise with a
complex spectrum should be observed provided the large enough
observation times are possible.
\end{itemize}
For both scenarios, the  $1/f$ noise spectrum should hold down to
arbitrary small (from practical point of view) frequencies.

Since the characteristic switching times for large aggregates can
be too long to be observable, at realistic noise frequencies  one
deals only with switchings of relatively small metastable
aggregates. To consider these aggregates as compact and
statistically independent one has to ensure that the aggregate's
switching between the metastable states does not cause
repopulation of the states outside the aggregate. Consequently,
the signs of their energies (counted from the chemical potential)
should not change. In addition, the Coulomb interactions with the
sites outside the aggregate should not affect the aggregate
dynamics. Both contributions are proportional to the
\textit{surface} of the aggregate, i.~e. $\propto N^{2/3}$.
Repeating the analysis of Sec.~\ref{s1} and ascribing the
probability factor $e^{-\gamma^*N^{2/3}}$ to allow for the surface
contribution one concludes that $S(\omega) \propto \omega^{-1}$
and $\bar{\alpha}=0$.

The controversial experimental results on the temperature
dependence of the flicker noise in the Coulomb glass, see e.~g.
Refs.~\onlinecite{exp},  can, in principle, be of fundamental
nature. Indeed, in the previous discussion we did not consider the
time fluctuations of the single-particle energies of the sites
forming the aggregates. These fluctuations are due to correlated
electronic hops, which can be more important for the noise than
for the stationary transport. An attempt to consider the role of
the fluctuations of site energies in hopping transport was made in
Ref.~\onlinecite{ours}. Somewhat later an important role of strong
fluctuations in the Coulomb glass was also emphasized in
Ref.~\onlinecite{Laikhtman}. We feel that correlated hopping may
indeed introduce a frequency cut-off to the noise spectrum; this
cut-off can be considered as a hallmark of correlated hopping.
This may be the case in the materials where experimental values of
$T_0$ are significantly smaller than those conventionally
expected.~\cite{Shklovskii-Efros}

Our conclusion regarding the absence of the  frequency lower
cut-off agrees favorably with available experimental results. The
temperature dependence is more problematic since different
temperature dependences were observed, see, e. g., analysis in
Ref.~\onlinecite{12}.

\acknowledgments This work was supported by the U. S. Department
of Energy Office of Science through contract No. W-31-109-ENG-38.
Prior to resubmission of the revised version of our manuscript
about the preprint by Burin and Shklovskii.~\cite{BS} We like to
thank A. L. Burin and B. I. Shklovskii for making their work
available before publication and for useful discussion.


\begin{thebibliography}{99}
\bibitem{endnote}Note that even if $\lambda > 1$, for some sites the inequality
$\delta V_i > \varphi_i$ can be violated. Thus, to be sure that
the chosen sites indeed form the aggregate obeying the criteria
mentioned above, one should impose some additional restriction on
the values of $\varphi$ for such ``improper" sites leading to some
statistical factor $\tilde \lambda$ for each of these sites. One
notes that the relative number $\beta$ of such sites depends on
the distribution functions of $\varphi_i$ and $V_i$ and does not
depend on $N$. As a result, the probability to form the aggregate
with $N$ sites is $\propto \tilde{\lambda}^{\beta N}$.
\end{thebibliography}
\end{document}